\documentclass[10pt,twocolumn,letterpaper]{article}
\pdfoutput=1

\usepackage{cvpr}
\usepackage{times}
\usepackage{epsfig}
\usepackage{graphicx}
\usepackage{amsmath}
\usepackage{amssymb}
\usepackage{subcaption}
\usepackage{cite}

\usepackage[breaklinks=true,bookmarks=false]{hyperref}

\cvprfinalcopy % *** Uncomment this line for the final submission

\def\cvprPaperID{****} % *** Enter the CVPR Paper ID here

\begin{document}

%%%%%%%%% TITLE
\title{Information Retrieval from the Digitized Books}

\author{Riya Gupta\\
IIIT Hyderabad\\
India\\
{\tt\small riyagupta771995@gmail.com}
% For a paper whose authors are all at the same institution,
% omit the following lines up until the closing ``}''.
% Additional authors and addresses can be added with ``\and'',
% just like the second author.
% To save space, use either the email address or home page, not both
\and
C.V. Jawahar\\
IIIT Hyderabad\\
India\\
{\tt\small jawahar@iiit.ac.in}
}

\maketitle
%\thispagestyle{empty}

%%%%%%%%% ABSTRACT
\begin{abstract}
Extracting the relevant information out of a large number of documents is a challenging and tedious task. The quality of results generated by the traditionally available full-text search engine and text-based image retrieval systems is not optimal. Information retrieval (IR) tasks become more challenging with the nontraditional language scripts, as in the case of Indic scripts. The authors have developed OCR (Optical Character Recognition) Search Engine to make an Information Retrieval \& Extraction (IRE) system that replicates the current state-of-the-art methods using the IRE and Natural Language Processing (NLP) techniques. Here we have presented the study of the methods used for performing search and retrieval tasks. The details of this system, along with the statistics of the dataset (source: National Digital Library of India or NDLI), is also presented. Additionally, the ideas to further explore and add value to research in IRE are also discussed.
\end{abstract}

%%%%%%%%% BODY TEXT
\section*{Keywords}
Optical Character Recognition, Information Retrieval \& Extraction, Metadata, Search Engine, Search Engine Optimization

 \section{Introduction}

Often the information retrieval systems designed to provide and assist the general populace with texts in different languages that are either not optimized or lack advanced language features, like getting similar results to a query. Even for the big digital database like Google Books, the contextual and semantic information is not available for the desired query. Short queries might not explain the full intent behind the search. However, suppose the Information Retrieval and Extraction systems (IRE) can be made capable in terms of obtaining semantically and contextually rich information from the ocean of data. In that case, it will improve the user experience. It will also help in the direction of extracting the required query. 

This work focuses on multimedia retrieval (image and text in Indic languages) in this IRE system, where the user enters a text query for which the system returns books that are indexed and ranked. Multilingual OCR\cite{7490115,DBLP:journals/corr/abs-1905-11739} is used to achieve this task. We know that optical Character Recognition (OCR) is the process of conversion of text or text-containing documents such as handwritten text, printed or scanned text images into a digital format that can be used for further processes. In this system, text recognition is performed on ancient Indic language documents. The final information retrieval system's goal is to demonstrate the high accuracy of the Indic OCR and the retrieval of the ancient texts in Indic languages. 

The challenges faced by IRE systems in Indic language scenarios are discussed in Section 2. Section 3 consists of the details and functionalities of the OCR Search Engine designed by the authors. Section 4 represents the ideas that can be implemented on top of the existing system.
%-------------------------------------------------------------------------

%-------------------------------------------------------------------------

\section{Practical Challenges associated with Indian IRE systems}

Modern state-of-the-art IRE systems suffer from various challenges for Indian data. These challenges can vary among different areas within information retrieval. These areas majorly include cross-lingual information retrieval (CLIR)\cite{inproceedings}, Web search, user modeling, filtering, classification, summarization, question answering, metasearch, semantic, and context-based search and extraction\cite{article}. Based on the current trends and research work, these are some of the challenges that the community faces at the moment:

%-------------------------------------------------------------------------
\subsection{Limited Research Aid}
\label{subsec:limited_aid}

According to the latest Indian Census, India has 24 officially recorded languages. These include Hindi and English. Although most of these languages have a similar grammatical structure and layout, the semantics and script differ drastically between them. Furthermore, a particular language varies widely in fonts, rendering schemes, dialects and other aspects depending on the region.

Moreover, the number of researchers contributing in the field of information retrieval and text-based challenges for Indian languages (Hindi, Bangla, Sanskrit, Tamil, Telugu and Malayalam) are comparatively smaller due to the diversity in them, as well as due to the involvement of various modalities for instance images, text or audio.

%-------------------------------------------------------------------------
\subsection{Limited amount of available data}

One of the reasons for the lack of research in building quality IRE systems for Indian languages is the lack of data for information retrieval. Even when the data is publicly available, it is not readily usable and requires extensive cleaning and preprocessing to make it suitable for the retrieval models. Current state-of-the-art methods like WordNet\cite{DBLP:journals/corr/abs-1807-05574}, ConceptNet\cite{conceptnet} are trained on large datasets for Roman and Latin scripts. Similar datasets are tedious to build due to the reasons stated in \ref{subsec:limited_aid}. Thus, the unavailability of quality data for this task hinders the progress in the field. 

Recently, several groups  from NDLI, IIT Kharagpur, have converted and digitalized a large number of historical documents like ”Bhagawadgitha”. These documents can act as a large dataset if annotated. However, even if it is present, the dataset is majorly noisy. Modern text recognition algorithms in this space are not designed to handle such noisy data and hence perform poorly. 

\begin{table}[h!]
 \begin{center}
 \begin{tabular}{ c    c }
 \includegraphics[width=0.45\linewidth, height=6cm]{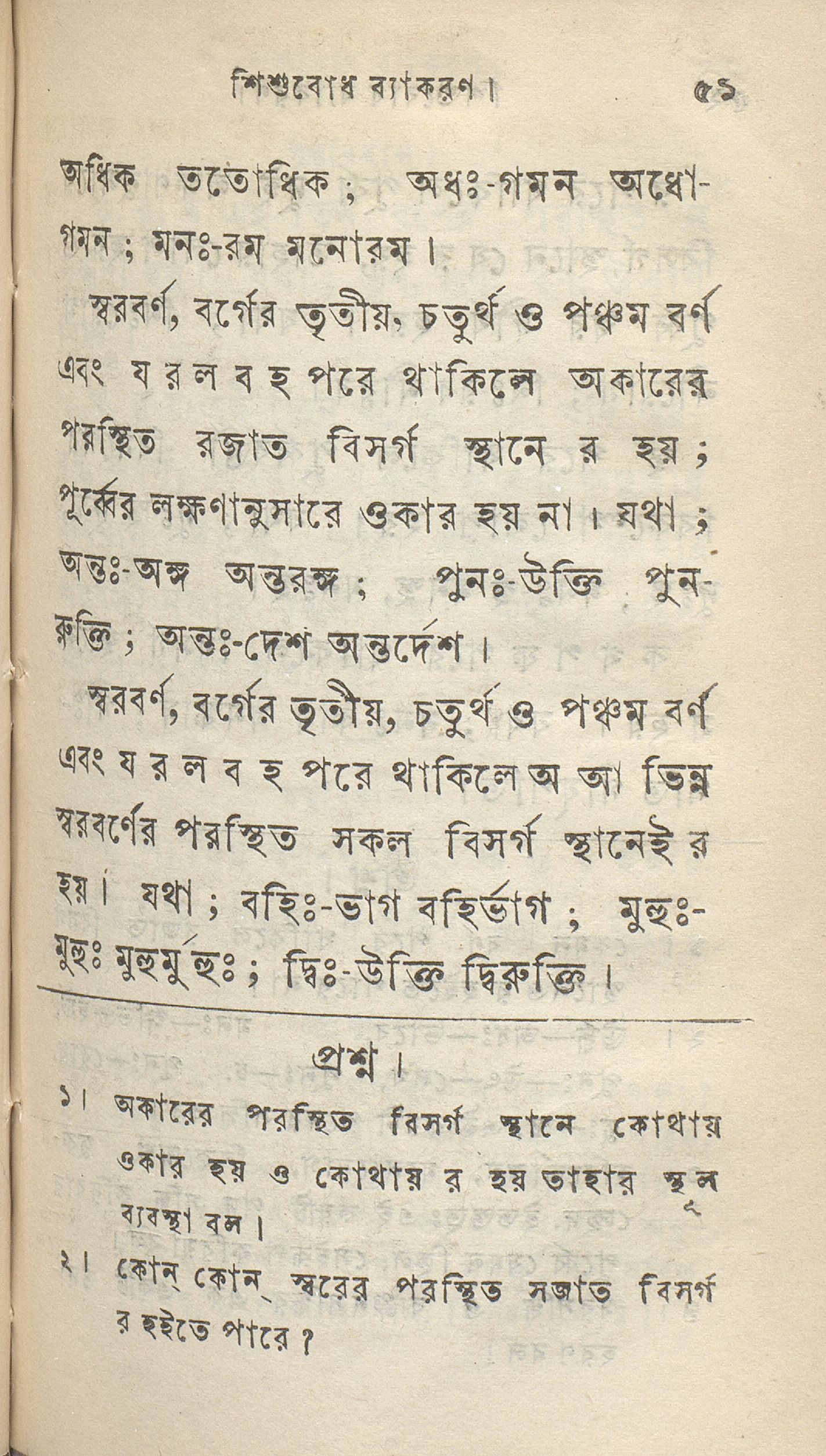}  &   \includegraphics[width=0.45\linewidth, height=6cm]{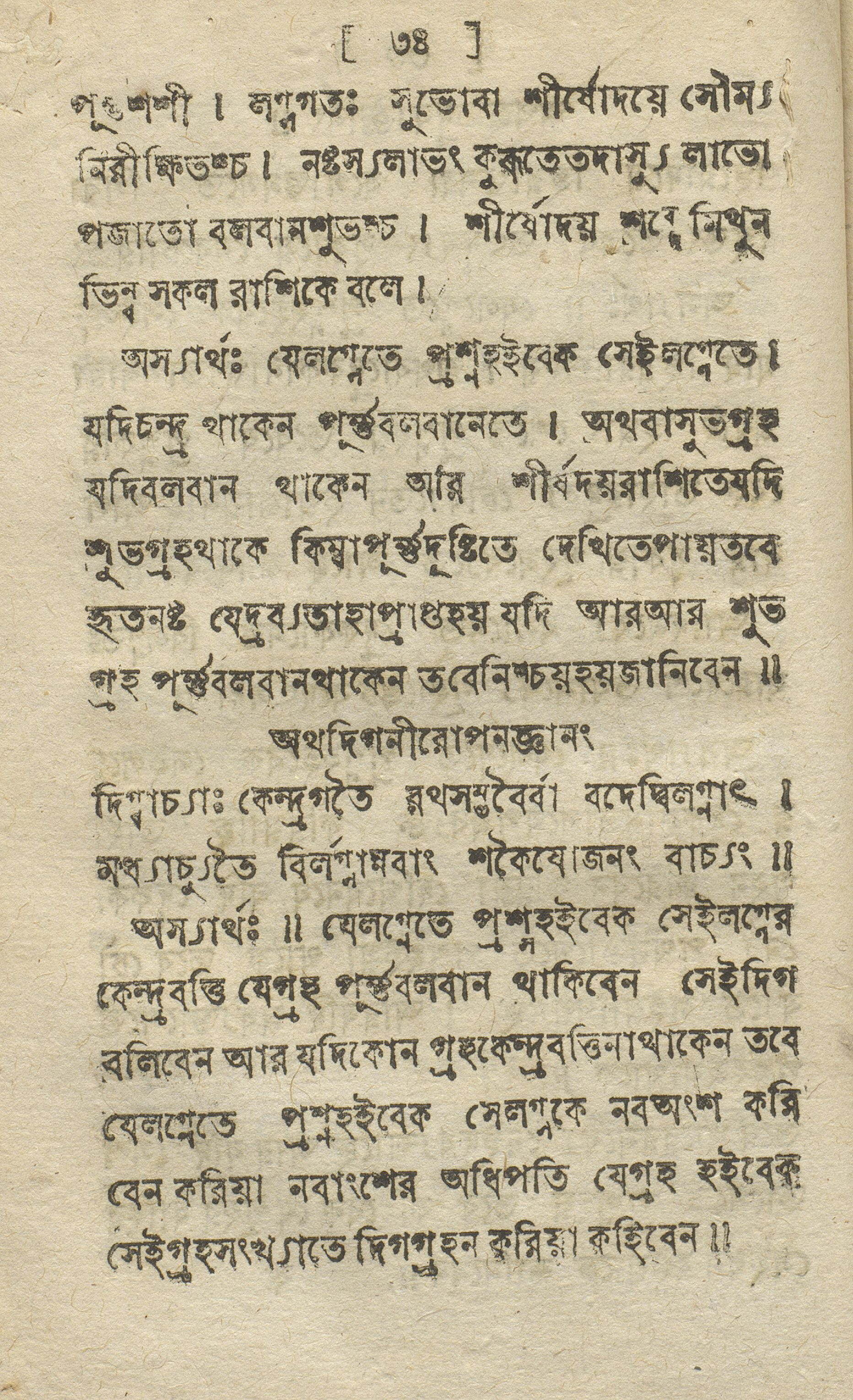}\\ 
 \end{tabular}

 \caption*{Figure 1: In this system, old and degraded document images are passed through an OCR, which hampers the quality of the generated text. The figure shows an example of a sample degraded document in Bangla language.}

\end{center}
\end{table}

Figure 1 shows a few examples of the data from the Bangla documents. The noise might include the warped lines within the documents, watermarked pages and traces of ink from the backside of the page.

%-------------------------------------------------------------------------
\subsection{Semantics and Context based Retrievals}

Web Search Engines and Information Retrieval are two aspects where web search is just a spectrum of Information retrieval. The information which is given as an input to the search bar does not always represent the practical and required information that a user demands. Understanding the rich semantic and context behind a text query is extremely important for the search engines to return relevant results. For instance, in the search engines Yahoo and Bing, if a person enters a search query 'Nokia', the results generated will be all the links containing Nokia, irrespective of their usability for the user. Whereas in Google search, the links are ranked in order of details about Nokia and currently available cellphones of Nokia showing a higher semantic understanding of the Google search engine than that of Yahoo and Bing. Similar features are needed to be implemented for the Indian language based IRE system. 

However, due to the complexity of Indian languages, newer techniques specifically designed for them are required for generating rich semantic embedding. Hence, building algorithms that can extract rich information from Indic texts and later combine them with user search history and search patterns can be beneficial.

%-------------------------------------------------------------------------

\section{What is OCR Search Engine?}

In recent years, there has been a lot of research in the field of text recognition, and efforts have been put into achieve an OCR that can perform well on Indian scripts. Indic scripts pose unique challenges for state-of-the-art OCR algorithms trained in English. For example, Sandhi rules and the difference in glyphs make it harder for the state-of-the-art OCR algorithms to work for such languages. makes it harder for the State-of-the-art OCR algorithms to work for such languages. Unavailability of IRE algorithms and dictionaries in Indic languages that can fetch similar words due to differences in the script and colossal amount of data to a given query hampers the search engine's performance. The developed OCR Search Engine that demonstrates the accuracy of the multilingual Indic OCR along with the retrieval accuracy.

This OCR Search Engine is an IRE system that uses Optical Character Recognition designed for Indian scripts to generate the text from the document images in the Indic languages. Currently, it supports Hindi, Tamil, and Telugu languages.

\begin{figure}[ht]
\begin{center}
%\fbox{\rule{0pt}{2in} \rule{0.9\linewidth}{0pt}}
   \includegraphics[width=1.1\linewidth]{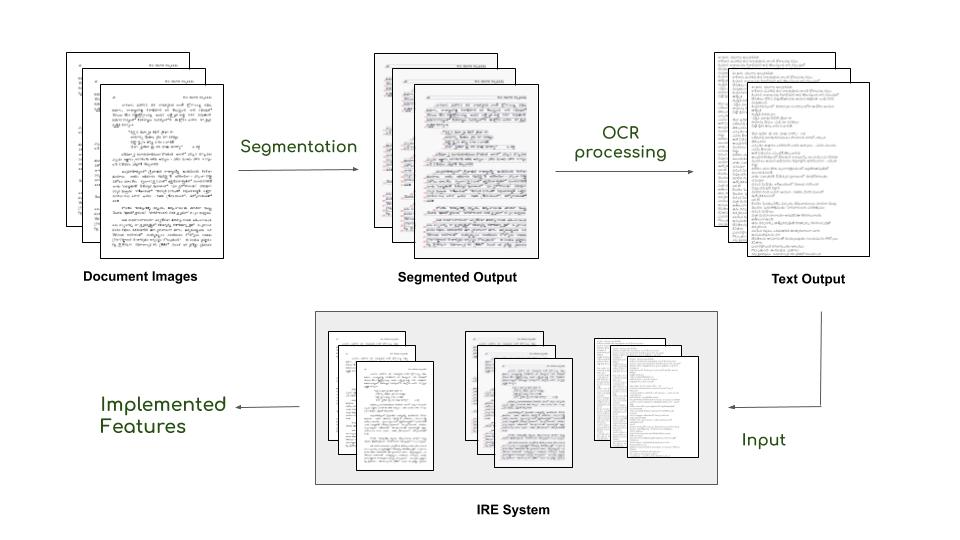}
\end{center}
   \caption*{Figure 2: Basic pipeline followed by the OCR Search Engine}
\label{fig:long}
\label{fig:onecol}
\end{figure}

The system is built using Python and PSQL, forming the back-end and HTML/CSS being used as the front end. It uses the text generated by the Indic OCR developed at IIIT Hyderabad, line-level segmentation, and document images to form the database of the system. The database is indexed using Elasticsearch, which is an open-source distributed Java package.

\subsection{Functionalities}

The IRE system designed by us performs retrieval tasks on multiple levels :

\begin{itemize}
\item \textbf{Query Level Search} 
\newline
The task is performed by entering the keyword/text in the search bar. There are multiple levels within the query level search like:
\begin{itemize}
\item Single Query Search: Search based on one keyword search
\item Multiple Query Search: Search based on more than one keywords
\item Exact Query Search : Search based on retrieving exact entered query (by using double quotes)
\item Query Processing: Removal of stopwords and implementation of tokenization and stemming.
\end{itemize}
\end{itemize}
\begin{itemize}
\item \textbf{Relevance and Visualization} 
    \begin{itemize}
    \item Relevance:  Based on the number of occurrence of a particular keyword in the database and number of hits for a particular book from users.
    \item Visualization: Highlighted query outputs within the resulting books using segmentation method. 
    \item Pages with figures and varying font styles and sizes retrievable.
    \end{itemize}
\end{itemize}   
\begin{itemize}
\item \textbf{Language Level Search} 
\newline
This level of search is completely based on the language chosen to retrieve the query.
\begin{itemize}
\item Transliteration:  A method of mapping from one system of writing to another based on phonetic similarity.
\item Single language search
\item Multilingual Search: Based on the search in more than one language and retrieval of multilingual books.
\end{itemize}
\end{itemize}
\begin{itemize}
\item \textbf{Search Engine Based} 
\begin{itemize}
\item Primary Filters: Filtering based on the languages (Hindi, Tamil and Telugu) and the other filters including content based filter (default), ISBN Code, Author name and Book title.
\item Secondary Filters: Filtering based on    Genre and Source provider of the book.
\end{itemize}
\end{itemize}

%-------------------------------------------------------------------------
\subsection{Dataset}

Table 1 shows the dataset and its details. Currently, the IRE system supports around 1.5 million documents. The detailed statistics are given in the following table. 
\begin{table}[!ht]
\begin{center}
\begin{tabular}{ |c|c|c|c| } 
 \hline
Language & Hindi & Tamil & Telugu \\ 
\hline
Books & 2287 & 2030 & 2574 \\ 
 \hline
Pages & 5,18,500 & 5,18,000 & 5,07,070 \\ 
 \hline
\end{tabular}
\vspace{0.3cm}
\caption{\label{tab:table-name}Statistics of the dataset}
\end{center}
\end{table}

% Table 1 shows the dataset and the details of the dataset. Currently, the IRE system supports around 1.5 million documents in 3 different languages i.e. Hindi, Tamil and Telugu. 
% \begin{table}[!ht]
% \begin{center}
% \begin{tabular}{ |c|c|c|c| } 
%  \hline
% Language & Hindi & Tamil & Telugu \\ 
% \hline
% Books & 2287 & 2030 & 2574 \\ 
% Pages & 5,18,500 & 5,18,000 & 5,07,070 \\ 
%  \hline
% \end{tabular}
% \vspace{0.5cm}
% \caption{\label{tab:table-name}Statistics of the dataset}
% \end{center}
% \end{table}

%-------------------------------------------------------------------------
\subsection{Design}
The OCR Search Engine is structured on the following basis : 
\begin{enumerate}
   \item \textbf{User End} : Home, Search page, View, See Results About. 
   \item \textbf{Server End}: Login, Database Status, Manual Updates to Database.
 \end{enumerate}
The following subsection represents the brief description of these ends.
 \subsubsection{User}
 
The user side is the client-side for retrieving the query entered by the user. The \textbf{Home} page is the launching page where the users can choose the filters for their search choices. This homepage presents the total number of documents that this IRE system supports. Users can select the filters, whether it be language (Hindi, Tamil, Telugu) or search criteria (title, ISBN, author, default: content). These will fetch the results based on the content present within the book and use text output generated from the Indic OCR.
 
\begin{figure}[ht]
\begin{center}
%\fbox{\rule{0pt}{2in} \rule{0.9\linewidth}{0pt}}
  \includegraphics[width=1.3\linewidth]{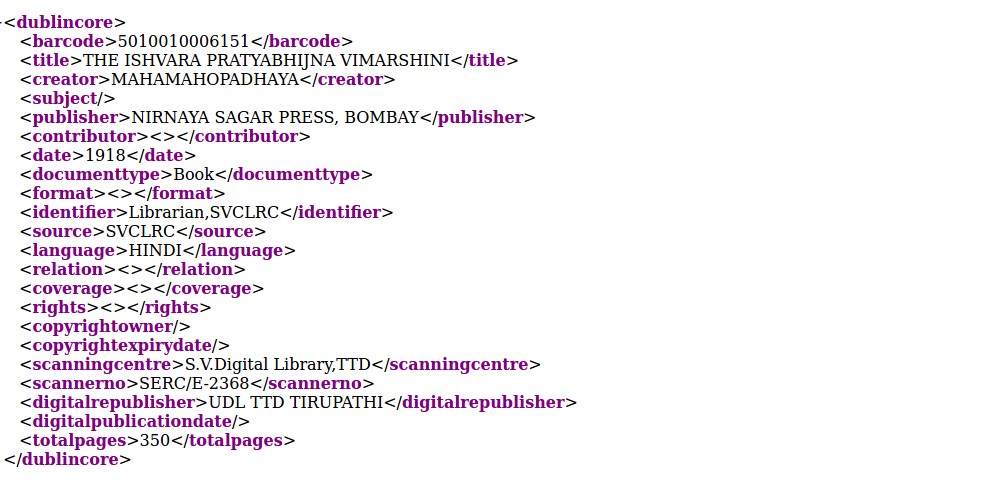}
\end{center}
   \caption*{Figure 3: Example of the metadata provided for one of the Hindi books}
\label{fig:long}
\label{fig:onecol}
\end{figure}
 
The other filters present are termed 'secondary filters' that include genre and the source if the book is extracted from the metadata provided. One of the examples of the metadata provided is given.

\begin{figure}[ht]
\begin{center}
%\fbox{\rule{0pt}{2in} \rule{0.9\linewidth}{0pt}}
   \includegraphics[width=1\linewidth]{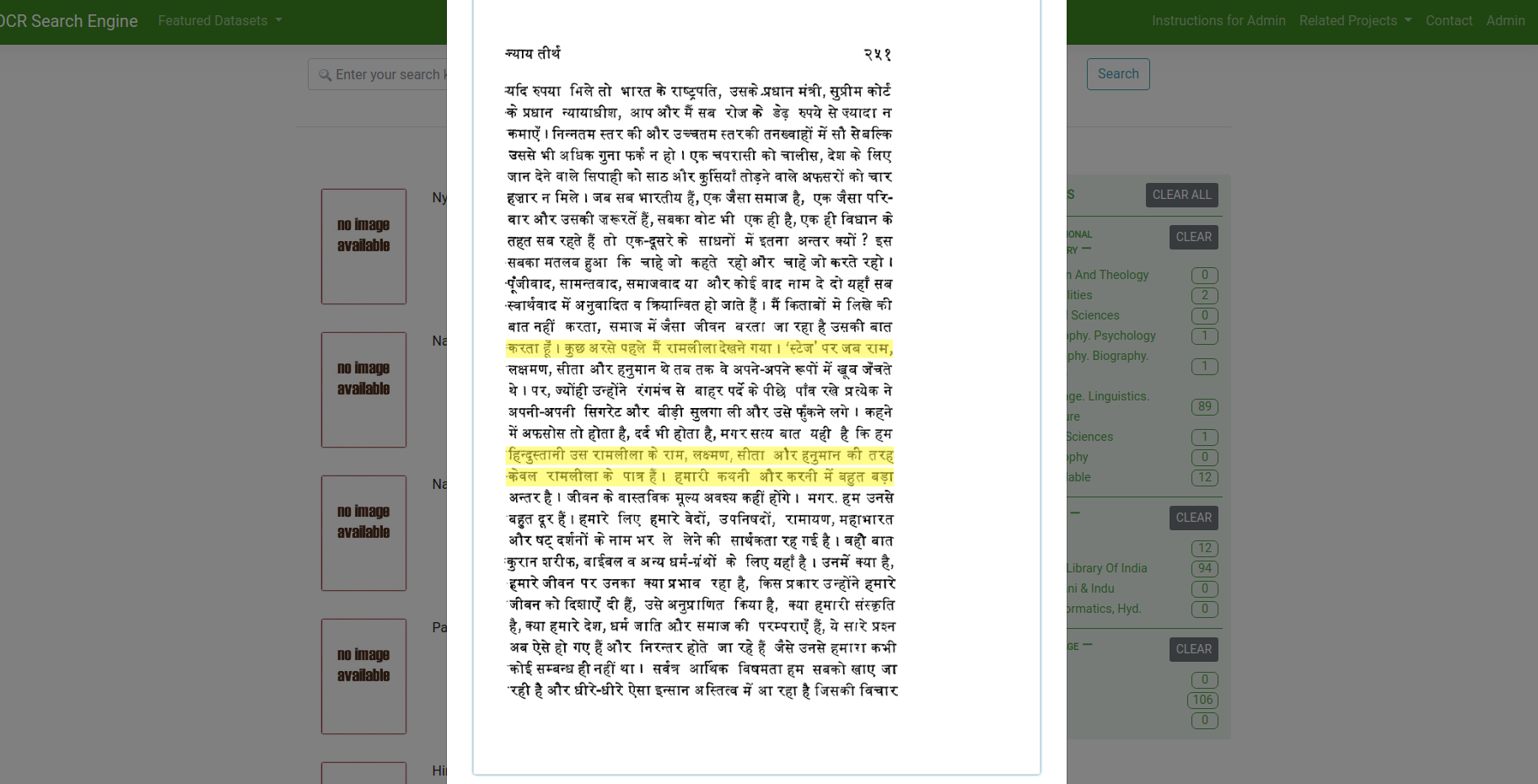}
\end{center}
   \caption*{Figure 4: Clicking 'See results about' opens up a pop up with highlighted lines in a document where the query is present}
\label{fig:long}
\label{fig:onecol}
\end{figure}

The search page displays all the retrieved results. Every result consists of the cover page of the book, author, title, language, ISBN, short abstract (if available), an option to \textbf{View} the book and \textbf{See results about}, clicking on which a pop-up appears that has all the pages of a book with the searched query present in them. As the current OCR system uses line-level segmentation, the line in which the query is present is highlighted for making it easier for the users to search it.

\begin{figure}[ht]
\begin{center}
%\fbox{\rule{0pt}{2in} \rule{0.9\linewidth}{0pt}}
   \includegraphics[width=1\linewidth]{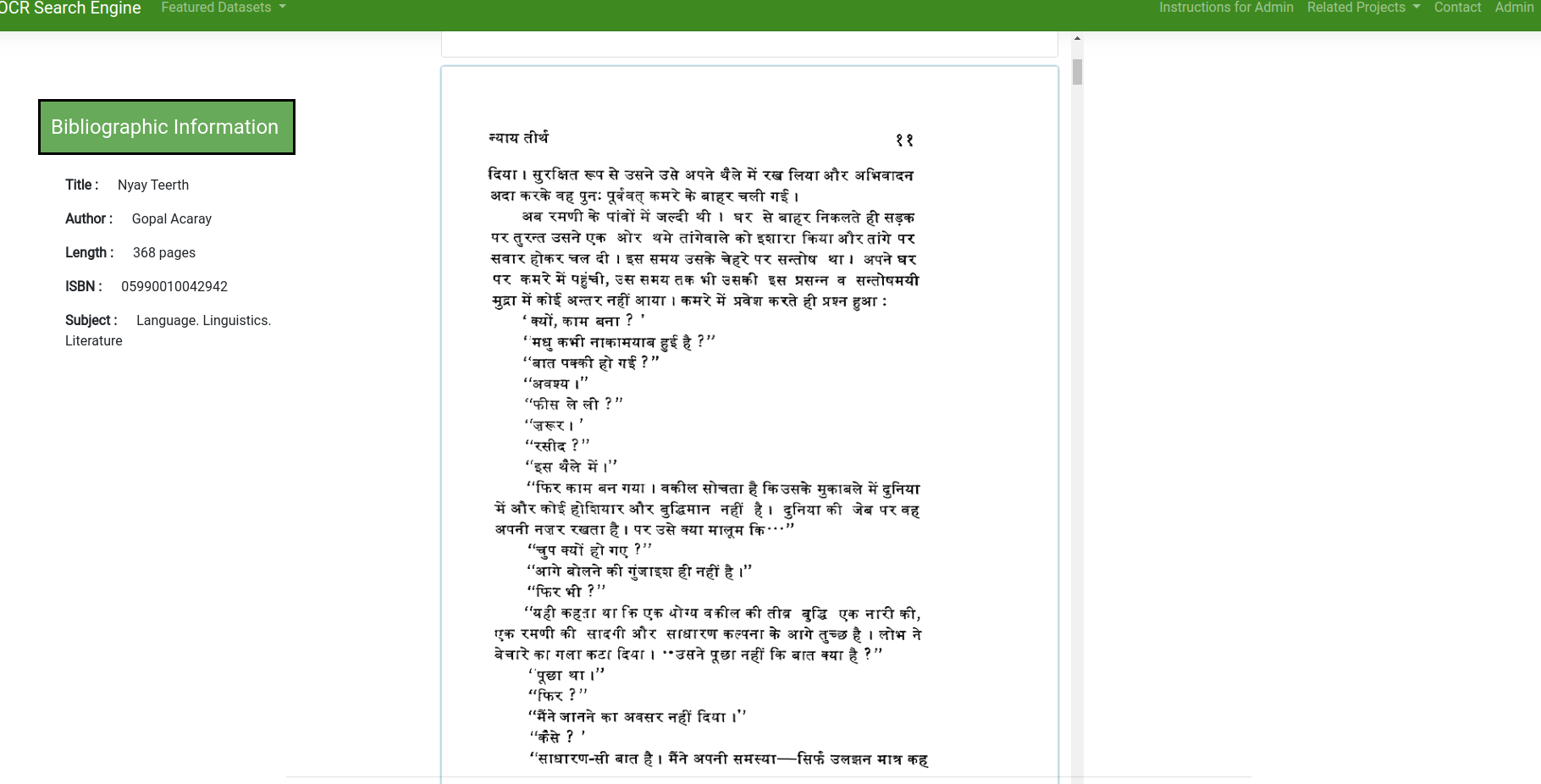}
\end{center}
   \caption*{Figure 5: Clicking "View" will open a new taskbar where users can view and read the entire book.}
\label{fig:long}
\label{fig:onecol}
\end{figure}

%-------------------------------------------------------------------------
\subsubsection{Server End}

The Admin can \textbf{Login} inside the IRE system and assign various roles to the other admins, whether they can alter the entries in the database or just monitor them. 
\textbf{Database Status} can be viewed as a graph that gives the total number of pages and the total number of books added within the entire database over a period. Featured datasets lists the datasets present within the system (currently by the National Digital Library Data). 

\textbf{Manual Updates} to the database are also possible. The Admin can edit/delete the books' details, including Abstract, ISBN and Author.

%-------------------------------------------------------------------------
\subsection{System Implementation}

The backend of the IRE system is built over the top of Django/python, and the frontend uses HTML/CSS and Javascript. The pages are indexed using Elasticsearch. The dynamic components of these IRE systems include the Search results, filters, visualization where the query searched is highlighted, admin authentication, and manually updating the database.

The admin/supervisor has the right to completely modify/delete/edit the dataset/database manually/automatically depending on his/her needs. This will not affect the already available dataset/database present in the system.

\textbf{\textit{Workflow}}: Users can view the retrieved results (retrieved books w.r.t. to the query searched), the highlighted queries, meta details, and the complete book(s). The download options are currently unavailable. The users do not need to log in to use the system, and this makes the search and the generated results accessible for everyone to view and read.

%-------------------------------------------------------------------------
\section{Future Work}

These are the few proposed ideas that can be implemented over the top of the existing system that has been built and discussed here.

%-------------------------------------------------------------------------
\subsubsection{Abstract Generation}

Abstracts are necessary when a user cannot dedicate his time reading the complete book. It allows a user to understand the summary of the book, thus enable a user to filter out books of his/her interest. This becomes more relevant in the case of Indic languages where there are religious texts in different languages; for instance, it might be the case that the user would not want to read the entire drama book in Hindi but the gist of a particular page. State-of-the-art attention-based models \cite{DBLP:journals/corr/WangL16,yin2018tranx,strzalkowski2000towards} can be used to generate a fluent, concise, and understandable summary of the complete book. This can even be used to create summaries of chapters or pages within the book, which will make the user experience more smooth and increase the system's usage. Relevant keywords, frequency within the document, and readability will contribute to the abstract generation. This feature can further be enhanced by adding an audio option, i.e., Text-To-Speech (TTS), which will recite the abstract of the book to the user. 

%-------------------------------------------------------------------------
\subsubsection{Semantic and Context Based IRE}

Usually, the searched queries are short or lack of intent behind the search. These queries might generate ambiguous results that the users do not expect. The current system designs fetch results using text processing techniques like tokenization and lemmatization in the Indic query entered or by extracting the exact matching query. However, it does not generate similar results or semantically rich results. For example, suppose the query 'Ram aur unke putra' is entered. In that case, the IRE system should be able to identify the semantics and relationship between the components 'Ram' and 'Putra' and generate relevant results containing details about 'Luv' and 'Kush'\cite{zamani2020generating,setchi2011semantic,DBLP:journals/corr/abs-1204-0186}.

By using natural language processing technologies, word similarity techniques, ontology techniques, etc., to understand the nature of the words and queries, the search results can be improved. Controlled vocabularies, dictionaries, thesaurus, and taxonomies are some of the basic ontology techniques that are used during the annotation. In text-based retrieval, these techniques improve the extraction of the information easily. The other method that can be used to achieve semantic Information retrieval is treating words as entities or concepts. In concept indexing, instances(entities) and abstract ideas(concepts) are identified within a text document and are linked to the ontological concepts. Methods like WordNet\cite{DBLP:journals/corr/abs-1807-05574}, OpenCyc\cite{inproceedings1} and ConceptNet\cite{10.1007/11880592_1} can be used to achieve some results in the case of Indic languages.

Additionally, the search results in the current system can be improved by using Query Expansion(QE)\cite{DBLP:journals/corr/abs-1708-00247}\cite{banchs2014query}. QE is performed by broadening the query by introducing additional keywords within the originally entered query. These usually include abbreviations, synonyms, antonyms, hyponyms, and hypernyms. Unsupervised or Supervised methods can be used to achieve this. Doc2Vec and Word2Vec can be used to capture the surface and semantic level similarity and perform Query Expansion. 

%-------------------------------------------------------------------------
\subsubsection{System Based}

Multiple features can be added to the system to enhance it. Machine Translation(MT)\cite{45610,sen-etal-2019-multilingual,choudhary-etal-2018-neural,das-etal-2016-study,DBLP:journals/corr/abs-1907-12437} is one  of the features that can be added. A user can type the query in one language and retrieve the results based on the meaning of that query in another language. For example, "Mahabharat" searched in "Hindi" should fetch "Mahabharatha" results from the languages Tamil and Telugu too. 

Image and Audio based search methods can also be added where the entered query can either be audio or image. The audio search method is Speech to Text conversion, where a user can speak, and the query will automatically be translated and entered into the search bar. Image-based/Document Information Retrieval\cite{395677,8489722} is another search method that can be implemented. In this work \cite{10.1007/978-3-319-16178-5_4} by VGG group, the retrieval is performed over the object category on British Library data.

Additionally, adding the most popular queries searched by the users over time and generation of related articles, magazines, as well as books concerning the book clicked by users are the few features that can be added over the top of the currently existing OCR Search Engine developed at CVIT, to make it more accessible and handy during usage.

%-------------------------------------------------------------------------
\subsection{Conclusion}
The end goal is to develop and deploy an enhanced Indic IRE system that can exploit the large corpora provided by the NDLI, and the output generated by the Indic OCR grew at CVIT. This system will improve the user search experience for the Indic languages and increase the popularity of reading digitized books amongst the general populace. This system is currently being developed by adding more functionalities and newer languages like Bangla and Sanskrit. 

{\small
\bibliographystyle{ieee_fullname}
\bibliography{egbib}
}

\end{document}